\documentclass{article}
\newcommand{\be}{\begin{equation}}
\newcommand{\ee}{\end{equation}}

\newcommand{\ndt}{\noindent}

\def\bea{\begin{eqnarray}}
\def\eea{\end{eqnarray}}
\def\beas{\begin{eqnarray*}}
\def\eeas{\end{eqnarray*}}
\def\sla{\raise.15ex\hbox{$/$}\kern-.57em}

\newcommand{\del}{{\partial}}

\newcommand{\half}{\frac{1}{2}}

\newcommand\fr[1]{\frac{1}{#1}}
\newcommand{\nn}{\nonumber}
\newcommand{\A}{{AdS$_{4}$}}

\usepackage{amsmath}
\usepackage{latexsym}
\usepackage{graphicx}
\usepackage{cite}

\begin{document}
\begin{titlepage}
\vskip 1cm
\centerline{\LARGE{\bf {Light-cone gravity in \A}}}

\vskip 1.5cm
\centerline{{Y. S. Akshay, Sudarshan Ananth and Mahendra Mali}} 
\vskip 1cm
\centerline{\em  Indian Institute of Science Education and Research}
\centerline{\em Pune 411008}
\vskip .5cm

\vskip 1.5cm

\centerline{\bf {Abstract}}
\vskip .5cm
\noindent We obtain a closed form expression for the Action describing pure gravity, in light-cone gauge, in a four-dimensional Anti de Sitter background. We perform a perturbative expansion of this closed form result to extract the cubic interaction vertex in this gauge.
\vfill

\end{titlepage}

\section{Introduction}
\vskip 0.3cm
\noindent Quantum field theories of gravity are plagued by divergences that seem to rule out any straightforward attempt to unite quantum theory and the general theory of relativity. There are however quite a few reasons to still study gravity as a quantum field theory. Foremost among these are the existence of surprising perturbative ties between gravity and the better understood Yang-Mills theory\footnote{\ndt The link between Gravity and Yang-Mills is surprising given the significant differences between the theories: dimensionful coupling versus dimensionless coupling, no color structure versus color traces and so on.}. These perturbative ties, stemming from the KLT relations~\cite{KLT}, tell us that tree level scattering amplitudes in gravity are the square of tree level scattering amplitudes in Yang-Mills theory. A Lagrangian (off-shell) origin for this relationship has also emerged~\cite{MHV} but a complete understanding of this important bridge between the two theories is still elusive.
\vskip 0.3cm
\ndt The KLT relations are valid on flat spacetime backgrounds so one question that motivates the present work is whether such perturbative ties between Yang-Mills and gravity survive when we move to curved spacetime backgrounds. It is not clear how a Yang-Mills $\leftrightarrow$ gravity relationship, on \A\ , would trace back to a stringy origin. However, it is interesting to study this connection from a purely field theoretical point of view.
\vskip 0.3cm
\ndt In this paper, we set up much of the light-cone (helicity) formalism essential to identifying such links at the level of the Action~\cite{MHV}. These perturbative ties seem to extend beyond the spin $1$-spin $2$ system. In particular, one may derive off-shell versions of these relations at cubic order for a spin $1$-spin $\lambda$ system~\cite{Ananth}. This is a further point of interest when examining the fate of these relations on curved backgrounds.
\vskip 0.3cm
\ndt Another motivation stems from our work on higher spin theories~\cite{Ananth}. There are various stumbling blocks when attempting to derive a Lagrangian describing an interacting higher spin theory. While the equations of motion are well studied~\cite{MV} we do not have an Action to quantize and it remains unclear whether we can define a consistent interacting S-matrix for such theories. The light-cone gauge approach to higher spin fields~\cite{BBB} yielded some of the first examples of consistent Lagrangians, on flat backgrounds, describing the cubic interactions of three fields, all of spin $\lambda$. The higher-spin story in curved backgrounds is different since the no-go theorems established for flat backgrounds no longer hold. To achieve a  light-cone formulation of higher spin fields ($\lambda>2$) on curved backgrounds, it is essential to have as a guidepost the Action for pure gravity ($\lambda=2$) on those backgrounds. This is one of the results we obtain here.

\vskip 0.3cm

\noindent In this paper, we describe how pure gravity is formulated in light-cone gauge on an \A\ background. This is achieved by making suitable gauge choices and using the constraint relations to eliminate the unphysical degrees of freedom. This will allow us to describe the Action of light-cone gravity on \A\ in a closed form purely using the physical degrees of freedom. We also perform a perturbative expansion of this gauge-fixed Action to first order in the gravitational coupling constant and comment on the resulting interaction vertex.

\vskip 0.3cm

\section{Preliminaries}
\label{sec:eingra}
\vskip 0.3cm

\noindent The Einstein-Hilbert action reads
\bea
\label{eh}
S_{EH}=\int\,{d^4}x\,{\mathcal L}\,=\,\frac{1}{2\,\kappa^2}\,\int\,{d^4}x\,{\sqrt {-g}}\,(\,{\mathcal R}\, - 2\,\Lambda\,)\ ,
\eea
\noindent where $g=\det{g_{\mu\nu}}$, $\mathcal R$ is the curvature scalar, $\Lambda$ is the cosmological constant of \A\ and $\kappa^2=8\pi G_N$ is the coupling constant in terms of the Newton constant. The gravity action on a manifold $M$ with boundary $\partial M$ can contain boundary terms, in addition to the bulk Einstein-Hilbert term. In general the form of the Action on such a background is 
\bea
\label{eh1}
S_{EH}=\int_{M} d^{4}x\, {\mathcal L}_{M}\, + \int_{\partial M}d^{3}x\, {\mathcal L}_{_{\partial M}}.
\eea
\ndt In this paper, we focus on the bulk term in (\ref {eh1}) which is sufficient to determine the equations of motion. It is important to note that one may always add boundary terms~\cite{GHY} to the Action that do not affect the equations of motion or Green functions\footnote{These added terms combine with surface terms generated by partial integrations of the bulk term.}.
\vskip 0.3cm
\noindent The light-cone gauge approach to formulating pure gravity in flat backgrounds has been studied in~\cite{BCL,SS,ABHS,lcg}. Here we formulate pure gravity in \A\ characterized by a cosmological constant $\Lambda$. As one would expect, this involves considerable modifications to the flat background results of~\cite{BCL} and we comment on these changes as and when they occur.
\vskip 0.5cm

\section{\A}

\ndt Consider a five-dimensional flat spacetime with metric $\eta_{_{MN}}\equiv(-1,1,1,1,-1)$ and co-ordinates $\xi^{M}$, $M=0\ldots4$. On this manifold, \A\ is defined as the four-dimensional hypersurface
\bea
\label{constraint}
-(\xi^{0})^{2}+(\xi^{1})^{2}+(\xi^{2})^{2}+(\xi^{3})^{2}-(\xi^{4})^{2}=R^2\ ,
\eea
with radius $R$. We now introduce local (Poincar\'{e}) co-ordinates $x^\mu\equiv(x^0,x^1,z,x^3)$ on \A\ 
\bea
\xi^{0}=\frac{R}{z}x^{0} \qquad\xi^{1}=\frac{R}{z}x^{1} \qquad\xi^{3}=\frac{R}{z}x^3\ ,\\
\xi^{2}=\frac{1}{2z}{\biggl [}R^{2}-\{-(x^{0})^{2}+(x^{1})^{2}+(x^{3})^{2}-z^{2}\}{\biggr ]}\ ,\\
 \xi^{4}=\frac{1}{2z}{\biggl [}R^{2}+\{-(x^{0})^{2}+(x^{1})^{2}+(x^{3})^{2}-z^{2}\}{\biggr ]}\ ,
\eea
which satisfy $(\ref{constraint})$. $z$ plays the role of a radial coordinate and divides the spacetime into two regions. We work here in the `patch' $z>0$ with $z=0$ being the \A\ boundary. The induced metric on this space is
\be
g_{\mu\nu}^{(0)}=\del_{\mu}\xi^{M}\del_{\nu}\xi^{N}\eta_{_{MN}}=\frac{R^{2}}{z^{2}}\eta_{\mu\nu}\ ,
\ee
\ndt where $\eta_{\mu\nu}$ is the usual Minkowski metric. We now switch to light cone co-ordinates $x^\mu\equiv(x^+,x^-,x^1,z)$ where
\bea
x^{\pm}=\frac{x^{0}\pm x^{3}}{\sqrt{2}}\ .
\eea
The cosmological constant for \A\ is
\bea
\label{cc}
\Lambda=-\frac{3}{R^2}\ .
\eea
\vskip 0.5cm

\section{Light-cone formulation}

\noindent Our aim is to study fluctuation on the \A\ background. The dynamical variable is the metric $g_{\mu\nu}$, which in the absence of all perturbations must reduce to $g_{\mu\nu}^{(0)}$. We work in light-cone gauge by making the following three gauge choices~\cite{BCL}
\bea
\label{lcg}
g_{--}\,=\,g_{-i}\,=\,0\quad ,\; i=1,z\ .
\eea
These choices are consistent with $g_{\mu\nu}^{(0)}$ since, in light-cone coordinates, $\eta_{--}=\eta_{-i}=0$. A fourth gauge choice will be made shortly. The metric is parametrized as follows
\bea
\label{gc}
\begin{split}
g_{+-}\,&=\,-\,e^\phi\ , \\
g_{i\,j}\,&=\,e^\psi\,\gamma_{ij}\ .
\end{split}
\eea
\noindent The fields $\phi\,,\,\psi$ are real while $\gamma_{ij}$ is a $2\times 2$ real, symmetric matrix. 
\vskip 0.3cm
\noindent The Euler-Lagrange equations corresponding to the Einstein-Hilbert Action read
\bea
{\mathcal R}_{\mu\nu}-\fr{2}g_{\mu\nu}\,{\mathcal R}\,=-\Lambda\,g_{\mu\nu}\ .
\eea
In light-cone gauge, a subset of the Euler-Lagrange equations which do not contain time derivatives ($\partial_+$) are treated as constraint equations. The first relevant constraint is ${\mathcal R}_{--}=0$ which reads
\bea
\label{con2}
2\,\partial_-\,\phi\,\partial_-\,\psi\,-\,2\,{\partial_-}^2\,\psi\,-\,{(\partial_-\,\psi)}^2\,+\,\frac{1}{2}\,\partial_-\,\gamma^{kl}\,\partial_-\,\gamma_{kl}\,=\,0\ .
\eea
\noindent A simple solution to this constraint relation may be obtained by making a fourth gauge choice
\bea
\label{4gc}
\phi\,=\,\frac{1}{2}\,\psi\ .
\eea
\noindent This allows us to solve equation (\ref {con2}) and obtain
\bea
\label{psi}
\psi=\frac{1}{4} \frac{1}{{\partial_-}^2} ( \partial_- \gamma^{i j} \partial_- \gamma_{i j}) + 2\, \ln\,\frac{R^2}{z^2}\ ,
\eea
with the $\fr{\partial_-}$ defined following the prescription in~\cite{SM}. The second term, in $\psi$, is essential to ensure that $g_{ij}$ and $g_{+-}$ reduce correctly to $g_{ij}^{(0)}$ and $g_{+-}^{(0)}$ respectively.
\vskip 0.3cm
\ndt In a flat background~\cite{BCL,SS} the solution to $\psi$ is simply $\psi_{\mbox {\footnotesize {flat}}}\!=\!\frac{1}{4} \frac{1}{{\partial_-}^2} ( \partial_- \gamma^{i j} \partial_- \gamma_{i j} )$ and the second term in (\ref {psi}) is absent.
\vskip 0.3cm
\ndt We now compute the determinant of $\gamma_{ij}$ from the second relation in (\ref{gc}) which implies that
\bea
\label{detg}
\det{g_{ij}^{(0)}}={\left(\frac{R^2}{z^2}\right)}^4\,\det{\gamma_{ij}^{(0)}}\ ,
\eea
with the $\{\,\}^{(0)}$ superscripts implying that all fluctuations are switched off. In this limit, the metric is simply $\frac{R^2}{z^2}$ times the Minkowski metric so the L.H.S of (\ref {detg}) is ${(\frac{R^2}{z^2})}^2$ thus implying that
\bea
\label{gamdet}
\det{\gamma_{ij}^{(0)}}={\left({\frac{z^2}{R^2}}\right)}^2\ .
\eea
\vskip 0.1cm
\ndt Note that in contrast to our result above, on a flat background, $\gamma_{ij}$ is unimodular~\cite{BCL,SS}. We choose the determinant of $\gamma_{ij}$ (which includes fluctuations) to be the same as in (\ref {gamdet}) - this is permitted since $\gamma_{ij}$ is a $2\times 2$ matrix that has only two physical degrees of freedom. This choice renders the fluctuation field, introduced in the next section, traceless making calculations easier. 
\vskip 0.3cm
\ndt The second constraint relation is ${\mathcal R}_{-i}=0$ which yields
\bea
\begin{split}
g^{-i}\,=\,-\mathrm{e}^{-\,\phi}\,\frac{1}{\partial_-}\bigg[\,&\,\gamma^{ij}\,\mathrm{e}^{\phi\,-\,2\,\psi}\,\frac{1}{\partial_-}\,{\Big \{}\,\mathrm{e}^{\psi}\,{\Big (}\,\frac{1}{2}\,\partial_-\,\gamma^{kl}\,\partial_j\,\gamma_{kl}\,-\,\partial_-\,\partial_j\,\phi \\
&\,-\,\partial_-\,\partial_j\,\psi\,+\,\partial_j\phi\,\partial_-\,\psi\,{\Big )}\,+\,\partial_l\,{\Big (}\,\mathrm{e}^{\psi}\,\gamma^{kl}\,\partial_-\,\gamma_{jk}\,{\Big )}\,{\Big \}}\,\bigg]\ .
\end{split}
\eea

\subsection{Light-cone Action}
\noindent The light-cone Action for gravity is
\bea
\label{lca}
S=\int d^3x\,\int dz\,{\mathcal L}=\frac{1}{2 \kappa^2}\int d^4x\, \sqrt{-g}\, \left( 2 g^{+-} {\mathcal R}_{+-} +g^{i j} {\mathcal R}_{i j} - 2 \Lambda\right)\ .
\eea
\ndt We now compute each term in the above expression, using the results listed thus far. We derive the following closed form expression for the Action in \A\ purely in terms of the physical degrees of freedom.
\bea
\label{aaction}
S\!\!\!\!\!\!\!\!&&=\frac{1}{2 \kappa^2}\int d^3x\,\int dz\,{\biggl \{}\; \frac{z^{2}}{R^{2}}e^{\psi}\left(2\del_{+}\del_{-}\phi + \del_{+}\del_{-}\psi - \half\del_{+}\gamma^{ij}\del_{-}\gamma_{ij}\right) \nonumber \\
&&-\frac{z^{2}}{R^{2}}e^{\phi}\gamma^{ij}\left(\del_{i}\del_{j}\phi + \half \del_{i}\phi\del_{j}\phi - \del_{i}\phi\del_{j}\psi - \frac{1}{4}\del_{i}\gamma^{kl}\del_{j}\gamma_{kl} + \half \del_{i}\gamma^{kl}\del_{k}\gamma_{jl}\right) \nn \\
&&-\frac{z^2}{2R^2} e^{\phi - 2\psi}\gamma^{ij}\frac{1}{\del_{-}}N_{i}\frac{1}{\del_{-}}N_{j} + \frac{2}{R^{2}}e^{\phi}\gamma^{zz}-2\,\frac{z^2}{R^2}\,e^\psi\,e^\phi\,\Lambda\,{\biggr \}}\ ,
\eea
with 
\beas
N_i&\!\!\!=&\!\!\!  \mathrm{e}^{\psi} 
\bigg( \frac{1}{2}\partial_- \gamma^{j k} \partial_i \gamma_{j k} -
  \partial_- \partial_i \phi -\partial_i
  \partial_- \psi + \partial_i \phi \partial_- \psi \bigg) \nonumber\\&&
  + \partial_k \bigg( \mathrm{e}^{\psi}\gamma^{j k} \partial_-
  \gamma_{ij} \bigg)\ .
  \label{req}
\eeas
Although $\phi=\fr{2}\psi$, we have not made this substitution in the result above - this makes  it easier to trace the origin, from (\ref {lca}), of each term in (\ref {aaction}). In obtaining the above result, we have dropped several boundary terms (see Section~\ref{sec:eingra}). 
\vskip 0.5cm
\ndt {\it {Deviations from flat spacetime results}}
\vskip 0.3cm

\ndt The three main differences between our result (\ref {aaction}) and the flat background Action in~\cite{BCL,SS} are the overall factor of $\frac{z^2}{R^2}$ in front of each line, the penultimate term proportional to $\gamma^{zz}$ and the last term, proportional to the cosmological constant. 

\section{Perturbative expansion}

In this section we obtain a perturbative expression, to cubic order in the fields, for the Action in (\ref {aaction}). We do this by making the following choice
\bea
\label{gammah}
\gamma_{i j}&=&\frac{z^2}{R^2}\left(\mathrm{e}^{H}\right)_{i j}\ , \nn \\ \nn\\
H&=&\begin{pmatrix} h_{11} & h_{1z}\\h_{1z} &-h_{zz}\end{pmatrix}\ ,
\eea
with $h_{zz}=-h_{11}$ as explained below equation (\ref {gamdet}). In terms of these fields, equation (\ref {psi}) reads
\bea
\label{newpsi}
\psi=-\fr{4}\frac{1}{{\partial_-}^2} 
\bigg[ \partial_- h_{ij} \partial_- h_{ij}\bigg]+\,2\,\ln\,\frac{R^2}{z^2}+{\it O}(h^4)\ .
\eea
In order to obtain a perturbative expansion of (\ref{aaction}) we simply use the results (\ref {gammah}) and (\ref {newpsi}).
\vskip 0.3cm
\ndt We now redefine 
\bea
h\,\rightarrow\,\fr{\sqrt 2\kappa}\,h\ .
\eea
\vskip 0.3cm
\ndt In terms of these fields, the Action at {\it {O}}$(h^2)$ is
\bea
S_2=\int d^3x \int dz\,{\mathcal L}_2\ ,
\eea
with
\beas
{\mathcal L}_2=+\frac{R^2}{2z^2}\partial_+h_{ij}\partial_-h_{ij}-\frac{R^2}{4z^2}\partial_ih_{kl}\partial_ih_{kl}-2\frac{R^2}{z^3}h_{ik}\partial_kh_{iz}+\frac{R^2}{z^4}h_{zk}h_{kz}\ .
\eeas
In the above we have made use of both (\ref {cc}) and the fact that $h_{kk}=0$. Notice from (\ref {aaction}) that the cosmological constant only appears in interaction vertices involving an even number of fields.
\vskip 0.3cm
\ndt At {\it {O}}$(h^3)$, the Action reads
\bea
S_3=\int d^3x \int dz\;\fr{\sqrt 2}\;{\mathcal L}_3\ ,
\eea
where
\beas
{\mathcal L}_3=&\!\!\!\!\kappa\,{\biggl\{} \!\!\!\!\!\!&-\fr{2}\frac{R^2}{z^2}\del_k h_{ik} \del_l h_{ij} h_{jl}+2\frac{R^2}{z^3}h_{iz}\del_k h_{ij} h_{jk}+\frac{R^2}{4z^2}h_{ij} \del_i h_{kl} \del_j h_{kl} \nn \\
&&-\frac{R^2}{2z^2} h_{ij} \del_i h_{kl} \del_k h_{jl}-2\frac{R^2}{z^3}h_{iz}\frac{1}{\del_-} (\del_- h_{lm}\del_i h_{lm}) \nn \\
&&+4\frac{R^2}{z^4} h_{iz} \frac{1}{\del_-}(h_{lz} \del_- h_{il})+4\frac{R^2}{z^3}\del_k h_{iz} \frac{1}{\del_-} (h_{lk} \del_- h_{il})\nn \\
&&+\fr{2}\frac{R^2}{z^2} \del_k h_{ik}\frac{1}{\del_-} (\del_- h_{lm}\del_i h_{lm})-4\frac{R^2}{z^3} \del_k h_{ik} \frac{1}{\del_-} (h_{lz} \del_- h_{il})\nn \\
&&+\frac{R^2}{z^2} \del_k h_{ik} \frac{\del_l}{\del_-}(h_{ml}\del_- h_{im})\nn \\
&&-2\frac{R^2}{z^2} h_{ij} \del_i \del_j B-6\frac{R^2}{z^3}h_{iz}\del_i B-4\frac{R^2}{z^4}h_{zz}B\,{\biggr \}}\ ,
\eeas
with
\bea
B=-\fr{8}\frac{1}{{\partial_-}^2} 
\bigg[ \partial_- h_{ij} \partial_- h_{ij}\bigg]\ .
\eea

\begin{center}
* ~ * ~ *
\end{center}

\ndt As expected, both the kinetic and cubic vertices in \A\ are far more involved than their flat background counterparts. It would be interesting to (1) extend our analysis to the quartic interaction vertex which is trickier because time derivatives start to appear and must be re-defined away as in appendix C of~\cite{ABHS}, (2) understand whether one can extract `amplitude-like' structures from these expressions as in~\cite{MHV}, (3) attempt a light-cone derivation of higher spin theories on \A\ following the methods of~\cite{BBB}, using the results presented here for guidance and (4) examine the formulation of pure gravity on other curved backgrounds,

\vskip 0.5cm
\noindent {\it {Acknowledgments}}
\vskip 0.3cm

\noindent We thank Hidehiko Shimada, Stefano Kovacs, Sunil Mukhi and Stefan Theisen for helpful discussions. This work is supported by the Max Planck Institute for Gravitational Physics through the Max Planck Partner Group in Quantum Field Theory and the Department of Science and Technology, Government of India, through the INSPIRE fellowship (YSA).

\newpage
\appendix 
\section{Useful results}

\beas
\Gamma_{++}^+&=&\frac{1}{2}{g^{+-}[2\partial_+g_{+-}-\partial_-g_{++}]}\\ [\baselineskip]
\Gamma_{+-}^+&=&0\\ [\baselineskip]
\Gamma_{--}^+&=&0\\ [\baselineskip]
\Gamma_{i-}^+&=&0\\ [\baselineskip]
\Gamma_{i+}^+&=&\frac{1}{2}g^{+-}[\partial_ig_{+-}-\partial_-g_{i+}]\\ [\baselineskip]
\Gamma_{ij}^+&=&-\frac{1}{2}g^{+-}\partial_-g_{ij}\\ [\baselineskip]
\Gamma_{--}^-&=&g^{+-}\partial_-g_{+-}\\ [\baselineskip]
\Gamma_{+-}^-&=&\frac{1}{2}\{g^{+-}\partial_-g_{++}+g^{-i}[\partial_-g_{i+}-\partial_ig_{+-}]\}\\ [\baselineskip]
\Gamma_{++}^-&=&\frac{1}{2}\{g^{+-}\partial_+g_{++}+g^{--}[2\partial_+g_{+-}-\partial_-g_{++}]\nonumber\\ 
&&+ g^{-i}[2\partial_+g_{i+}-\partial_ig_{++}]\}\\ [\baselineskip]
\Gamma_{+i}^-&=&\frac{1}{2}\{g^{+-}\partial_ig_{++}+g^{--}[\partial_ig_{+-}-\partial_-g_{i+}]\nonumber\\ [\baselineskip]
&&+g^{-j}[\partial_ig_{+j}+\partial_+g_{ij}-\partial_jg_{+i}]\}\\ [\baselineskip]
\Gamma_{-i}^-&=&\frac{1}{2}\{g^{+-}[\partial_ig_{+-}+\partial_-g_{+i}]+g^{-j}\partial_-g_{ij}\}\\ [\baselineskip]
\Gamma_{ij}^-&=&\frac{1}{2}\{g^{+-}[\partial_ig_{+j}+\partial_jg_{+i}-\partial_+g_{ij}]-g^{--}\partial_-g_{ij}\nonumber\\
&&+g^{-k}[\partial_ig_{kj}+\partial_jg_{ik}-\partial_kg_{ij}]\}\\ [\baselineskip]
\Gamma_{jk}^i&=&\frac{1}{2}\{-g^{-i}\partial_-g_{jk}+g^{im}[\partial_jg_{mk}+\partial_kg_{mj}-\partial_mg_{jk}]\}\\ [\baselineskip]
\Gamma_{-j}^i&=&\frac{1}{2}g^{ik}\partial_-g_{kj}
\eeas
\beas
\Gamma_{+-}^i&=&\frac{1}{2}g^{ij}[\partial_-g_{j+}-\partial_jg_{+-}]\\ [\baselineskip]
\Gamma_{+j}^i&=&\frac{1}{2}\{g^{-i}[\partial_jg_{+-}-\partial_-g_{+j}]+g^{ik}[\partial_jg_{+k}+\partial_+g_{kj}-\partial_kg_{+j}]\}\\ [\baselineskip]
\Gamma_{++}^i&=&\frac{1}{2}\{g^{-i}[2\partial_+g_{+-}-\partial_-g_{++}]+g^{ij}[2\partial_+g_{+j}-\partial_jg_{++}]\}\\ [\baselineskip]
\Gamma_{--}^i&=&0\\ [\baselineskip]
\Gamma_{ij}^j&=&\frac{1}{2}\{-g^{-j}\partial_-g_{ij}+g^{jl}[\partial_jg_{li}+\partial_ig_{lj}-\partial_lg_{ij}]\}\\ [\baselineskip]
\, \\
\gamma^{ij}&=&\frac{R^{2}}{z^{2}}(e^{-H})_{ij} \\
\gamma^{ij}\gamma_{ij}&=&2 \\ 
\gamma^{ij}\del_{k}\gamma_{ij}&=&\frac{4}{z}\delta_{kz} \\
\, \\
g^{\mu\nu}g_{\mu\rho}&=&\delta^{\nu}_{\rho}\implies g^{++} = g^{+i} = 0 \\
g_{+i}&=&-g_{+-}g_{ij}g^{-j} \\
g_{++}&=&-e^{\psi}g^{--}+e^{\phi}g^{-i}g_{+i} \\ 
g^{+-}&=&-e^{-\phi}\nn \\ 
g^{ij}&=& e^{-\psi}\gamma^{ij} \\ 
\,\\
\sqrt{-g}&=&\frac{z^{2}}{R^{2}}e^{\psi}e^{\phi} \\ 
\,\\
\,\\
\eeas

\end{document}